\documentclass{elsart}
\usepackage {graphicx,amsmath}
\begin{document}
\begin{frontmatter}
\title{Limited validity of West and Yennie integral
formula for elastic scattering of hadrons}
\vspace*{-0.3cm}
\author{Vojt\v{e}ch Kundr\'{a}t, 
Milo\v{s} Lokaj\'{\i}\v{c}ek}
\footnote{e-mail: kundrat@fzu.cz,lokaj@fzu.cz}
\address{Institute of Physics of the
AS CR, v. v. i. 182 21 Prague 8, Czech Republic}
\author {Ivo Vrko\v{c}}
\footnote{e-mail:vrkoc@math.cas.cz} 
\address{Institute of Mathematics of the AS CR, v.v.i.
115 67 Prague 1, Czech Republic}
\maketitle
\begin{abstract}
The commonly used West and Yennie 
integral formula for the relative phase between
the Coulomb and elastic hadronic amplitudes might be 
consistently applied to only if the hadronic amplitude 
had the constant ratio of the real to the imaginary parts 
at all kinematically allowed values of four momentum 
transfer squared. 
\end{abstract}  
\end{frontmatter}

In our recent paper \cite{kun1} we have pointed out
that the integral formula of West and Yennie \cite{west}  
describing the relative phase between the Coulomb and hadronic 
high-energy elastic scattering amplitudes may be applied 
only to the elastic hadronic amplitudes $F^N(s,t)$
having the constant ratio between real and imaginary
parts at all values of the four momentum transfer 
squared $t$; $s$ being the value of the total CMS energy 
squared. As the given statement has not been explicitly
proved it seemed for a series of colleagues as unreasoned. 
The corresponding reasoning will be, therefore, given in 
the following.

West and Yennie \cite{west} derived for 
the phase function $\alpha \Phi (s,t)$ 
in the case of charged point-like nucleons 
($s \gg m^2$, $m$ being nucleon mass) the 
formula
\begin{equation}
\alpha \Phi (s,t) = \mp \alpha \bigg [ \ln \bigg ( {{-t} \over s} \bigg ) -
\int_{-4 p^2}^{0} {{ d \tau} \over {|t - \tau|}} \bigg ( 
1 - {{F^N(s,\tau)} \over {F^N(s,t)}} \bigg ) \bigg ]
\label{wy1}
\end{equation}
where $p$ is the value of the momentum in CMS
and $\alpha = 1/137.036$ is the fine structure constant.
The phase function $\alpha \Phi (s,t)$ is regarded always
to be real, which requires to hold for any admissible $t$: 
 \begin{equation}
\int_{-4 p^2}^{0} {{ d \tau} \over {|t - \tau|}} \Im \bigg ( 
{{F^N(s,\tau)} \over {F^N(s,t)}} \bigg ) \; \equiv \; 0.
\label{wy2}
\end{equation}
The condition (\ref{wy2}) may be transformed
to the condition
\begin{equation}
I(s,t) = \!\!\!\! \int\limits_{-4 p^2}^{0} {{d \tau} \over {|t - \tau|}}
\bigl [ \Re F^N(s,t) \Im F^N(s,\tau) - \Re F^N(s,\tau) \Im F^N(s,t) \bigr ]
\; \equiv \; 0.
\label{wy3}
\end{equation}
Introducing the phase $\zeta(t)$ and 
the modulus $F(t)$ of the complex hadronic 
amplitude (the dependence on the fixed $s$ 
being depressed in the following) by
\begin{equation}
F^{N}(s,t) \; = \; i F(t) e^ {-i \zeta (t)},
\label{nu1}
\end{equation}
it is possible to write
\begin{equation}
I_1(t) \; \equiv \; I_2(t)  
\label{wy4}
\end{equation}
where
\begin{equation}
I_1(t) = \int\limits_{-4 p^2}^{t} d \tau f(t,\tau),
\hspace*{0.5cm}
I_2(t) = \int\limits_{t}^{0} d \tau f(t,\tau) 
\label{df1}
\end{equation}
and 
\begin{equation}
f(t, \tau) = \begin{cases}
{{\sin[\zeta(t) - \zeta(\tau)]} \over{t - \tau}} F(\tau) &
\text{for $\tau \ne t$},\\
[\zeta(\tau)]' F(\tau)& \text{for $\tau = t$}
\end{cases}
\label{fu1}
\end{equation}
where the factor $\sin[\zeta(t) - \zeta(\tau)]/(t - \tau)$
is symmetrical in both the variables $t, \tau \in [-4p^2, 0]$. 
The function $f(t,\tau)$ is continuous 
and bounded if $\zeta(t)$ is continuous and its
derivatives are bounded for any $t \in [-4p^2, 0]$.
Similar properties may be assumed for the modulus 
$F(t)$ that is non-zero in the whole interval 
$[-4p^2, 0]$ with the only exception at
$t = - 4 p^2$.
It holds
\begin{equation}
\lim_{\tau \rightarrow t}
 {{ \sin [\zeta(t) - \zeta(\tau) ]} \over {t-\tau}} = 
[\zeta(t)]'
\label{nu2}
\end{equation}
and both $I_1(t)$ and $I_2(t)$ are proper
integrals \cite{vebl}. 

It is then possible to write 
\vspace*{-0.4cm} 
\begin{eqnarray}
\hspace*{0.5cm}[I_1 (t)]' &=& \int\limits_{-4 p^2}^{t} \!\!d\tau \;
{{\partial} \over {\partial t}}  f(t,\tau) + f(t,t) \; = \;
\int\limits_{-4 p^2}^{t} \!\! d\tau \; g(t,\tau) + f(t,t),
\nonumber\\
\hspace*{0.5cm}[I_2 (t)]' &=& \;\int\limits_{t}^{0} d\tau \;
{{\partial} \over {\partial t}} f(t,\tau) - f(t,t) \;=\;
\int\limits_{t}^{0} d\tau \; g(t,\tau) - f(t,t)
\label{nu3}
\end{eqnarray}
where
\vspace*{-0.5cm}
\begin{eqnarray}
g(t,\tau) &=& {{\partial} \over {\partial t}} 
f(t,\tau) =
\nonumber\\
 &=& \begin{cases}
 {{\cos [\zeta (t) - \zeta (\tau)]
[\zeta(t)]' (t-\tau) - \sin[\zeta (t) - \zeta (\tau)]}
\over {(t-\tau)^2}} F(\tau) & \text{for $t \ne \tau$}, \\
{1 \over 2} [\zeta(t)]'' F(t) & \text{for $t=\tau$}.
\end{cases}
\label{nu4}
\end{eqnarray}
Eq.(\ref{wy4}) passes now to the form
\begin{equation}
\int\limits_{-4 p^2}^{t} \!\! d\tau \; g(t,\tau) -
\int\limits_{t}^{0} d\tau \; g(t,\tau) + 2 f(t,t) \; \equiv \; 0
\label{fu8}
\end{equation}
which holds for each $t \in [-4p^2, 0]$.
Both the integrals in Eq. (\ref{fu8}) are 
proper integrals similarly as in Eq. (\ref{wy4}) 
(due to the assumed finite value of $\zeta (t)''$ -
see Eq.(\ref{nu4})). All higher derivatives 
of $I_1(t)$ and $I_2(t)$ (if they exist)
can be derived in a similar way. It is evident that 
they are continuous and bounded, too. From Eq. (\ref{wy4})
it can be easily derived that they should fulfill similar 
condition, i.e.,
\begin{equation}
I_1^{(n)}(t) \; \equiv \; I_2^{(n)}(t). 
\label{wy5}
\end{equation}
It may be shown that all equations 
(\ref{wy4}), (\ref{fu8}) and (\ref{wy5})
are fulfilled if
\begin{equation}
\zeta(t) \; = \; \zeta(\tau) \; \equiv \; const. 
\label{nu5}
\end{equation}
And we should ask whether this is the unique 
solution of the given problem. This can be 
answered with the help of the following theorem.

Theorem: {\it{ Let $\zeta(t)$ be continuous 
function on the closed interval $J = [-a,0]$, 
$a>0$; let $F(t)$ be continuous function and 
nonzero with exception of end points defined 
also on $J$. Suppose that for all $t$ and 
$\tau$ from $J$ it holds}}
\begin{equation}
\max_{t} \; \zeta(t) - \zeta(\tau) < \pi. 
\label{nu6}
\end{equation}
{\it{If for each \; $t \in J$ }} 
\begin{equation}
\int\limits_{-a}^{t} d \tau \;
{{\sin[\zeta(t) - \zeta(\tau)]} \over{t - \tau}} \;  F(\tau)
- \int\limits_{t}^{0} d \tau \;
{{\sin[\zeta(t) - \zeta(\tau)]}\over{t - \tau}} \;  F(\tau)
\; \equiv \; 0,
\label{nu7}
\end{equation}
{\it{then the function $\zeta(t)$ is a constant 
function on $J$}}.

Proof: Let us assume that the function $\zeta$ is 
not constant. Let us define $t_{max}$ as 
$\zeta(t_{max}) = \max_{t} \zeta (t)$. If there are
more such points we can take any of them. Let $M$
be the set of all the numbers from $J$ such that
$\zeta(t) < \zeta(t_{max})$. Owing to the assumption that
the function $\zeta$ is continuous and non-constant 
the Lebesgue measure of the set $M$, i.e. $\mu(M)$,
is positive.

Let us define further
\begin{equation}
L_1(t) = \int\limits_{-a}^{t} d \tau \;
{{\sin[\zeta(t) - \zeta(\tau)]} \over{t - \tau}} \;  F(\tau) 
\nonumber
\end{equation}
and 
\begin{equation}
L_2(t) = \int\limits_{t}^{0} d \tau \;
{{\sin[\zeta(t) - \zeta(\tau)]}\over{t - \tau}} \;  F(\tau).
\nonumber
\end{equation}
Let us introduce now $L(t_{max}) = L_1(t_{max})-L_2(t_{max})$.
For $-a \le \tau \le t_{max}$ it holds 
$\zeta (t_{max}) - \zeta (\tau) \ge 0$ and due to
the validity of condition (\ref{nu6}) we obtain
$\sin [ \zeta(t_{max}) - \zeta(\tau)] \ge 0$; 
as \; $t_{max} - \tau \ge 0$ and we can assume
the function $F(\tau)$ to be positive it holds in 
the corresponding interval $L_1(t_{max}) \ge 0$.

Similarly also for $L_2(t_{max})$: for
$t_{max} \le \tau \le 0$ it holds 
$\zeta(t_{max}) - \zeta(\tau)) \ge 0$. 
Due to (\ref{nu6}) it holds also
$\sin[ \zeta(t_{max}) - \zeta (\tau)] \ge 0$.
Owing to \; $t_{max} - \tau \le 0$, one obtains
\begin{equation}
{{\sin[\zeta(t_{max}) - \zeta(\tau)]}\over{t_{max} - \tau}} \le 0
\nonumber
\end{equation}
and therefore $L_2(t_{max}) \le 0$. Consequently
it holds $L(t_{max}) \ge 0$ and as the set $M$
has a positive measure we obtain $L(t_{max}) > 0$,
which contradicts the requirement (\ref{nu7}).
And the theorem is proved.
 
It may be easily seen that condition 
(\ref{nu6}) is fulfilled by the
$t$ dependence of the hadronic phases
used in all current phenomenological 
models of elastic hadron scattering and
if we put the lower integral limit 
$-a=-4p^2$ then the Theorem corresponds 
fully to our problem.

Thus the function
\begin{equation}
\zeta(t) \; = \; const    
\label{nu8}
\end{equation}
represents the only solution of Eq. (\ref{wy4}).
Therefore  Eq. (\ref{nu8}) represents the unique 
possibility for the $t$ dependence of  
$\zeta(t)$ function, if the relative phase 
between the Coulomb and hadronic amplitudes 
(given by integral formula of West and Yennie
(\ref{wy1})) is to be real quantity.

The problem of the $t$ independence of hadronic phase 
was mentioned for the first time in Ref. \cite{amal}.
As the authors felt a kind of uncertainty, they used 
this independence as an assumption for small values of 
$|t|$. However, our result has shown now that the 
integral West and Yennie formula for the relative phase
between the Coulomb and elastic hadronic amplitudes 
may be used in a consistent way only for the elastic 
hadronic amplitudes having $t$ independent hadronic 
phase. The contemporary experimental data on high-energy 
elastic nucleon scattering show, however, convincingly
that the ratio $\rho$ of real and imaginary parts must 
be $t$ dependent. Consequently, a correct approach 
that may be used for the description of interference 
between the Coulomb and hadronic scattering at present 
seems to be the eikonal model (see, e.g., \cite{kun1,kun2}),
only.

Similar conclusions seem to be valid also for the $pp$ elastic
scattering at the energy of 14 TeV at the LHC. The analyzed 
phenomenological model predictions \cite{tdr} clearly show
that the region of momentum transfers where the quantity 
$\rho (t)$ may be considered as being approximately 
constant (i.e., $t$ independent) is much narrower 
then that corresponding to contemporary measured energies. 
Thus the use of simplified West and Yennie formula for
the total elastic scattering amplitude cannot be 
justified in this energy region, too.

\end{document}